\documentclass[aps,prl,amsmath,amssymb,twocolumn,
nofootinbib,showpacs,tightenlines,superscriptaddress]{revtex4}

\usepackage{epsfig,graphicx}

\usepackage{bm}
\newcommand{\be}{\begin{equation}}
\newcommand{\ee}{\end{equation}}
\newcommand{\bea}{\begin{eqnarray}}
\newcommand{\eea}{\end{eqnarray}}
\newcommand{\nn}{\nonumber}

\begin{document}
\title{How Does Casimir Energy Fall?}
\author{Stephen A. Fulling}
\email{fulling@math.tamu.edu}
\homepage{www.math.tamu.edu/~fulling/}
\affiliation{Departments of Mathematics and Physics, Texas A\&M University,
College Station, TX 77843-3368 USA}

\author{Kimball A. Milton}
\email{milton@nhn.ou.edu}
\homepage{www.nhn.ou.edu/~milton/}
\author{Prachi Parashar}
\affiliation{Oklahoma Center for High-Energy Physics and Homer L. Dodge 
Department of Physics and Astronomy, University of Oklahoma,
Norman, OK 73019-2061 USA}

\author{August Romeo}
\affiliation{Societat Catalana de F\'isica, Laboratori 
de F\'isica Matem\`atica (SCF-IEC), 08028 Barcelona, 
Catalonia, Spain}
\author{K.V. Shajesh}
\author{Jef Wagner}
\affiliation{Oklahoma Center for High-Energy Physics and Homer L. Dodge 
Department of Physics and Astronomy, University of Oklahoma,
Norman, OK 73019-2061 USA}
%%%%%%%%%%%

%%%%%%%%%%%

\date{\today}


\begin{abstract}
Doubt continues to linger over the reality of quantum vacuum energy.
There is some question whether fluctuating fields gravitate at all, or do so
anomalously. Here we show that for the
simple case of parallel conducting plates, the associated Casimir energy
gravitates just as required by the equivalence principle, and that therefore
the inertial and gravitational masses of a system possessing Casimir energy
$E_c$ are both $E_c/c^2$.  This simple result disproves
recent claims in the literature.
We clarify some pitfalls in the calculation that can lead to spurious 
dependences on coordinate system.
\end{abstract}

\maketitle

The subject of quantum vacuum energy (the Casimir effect) dates from the
same year as the discovery of renormalized quantum electrodynamics, 1948
\cite{casimir}.
It puts the lie to the naive presumption that zero-point energy is not
observable.  
On the other hand, 
it continues to be surrounded by controversy, 
in large part because
sharp boundaries give rise to divergences in the local energy density near
the surface (see Refs.~\cite
{Deutsch:1978sc,Candelas:1981qw,Graham:2003ib}).
 The most troubling aspect of these divergences is in the 
coupling to gravity.  Gravity has its source in 
the local energy-momentum tensor, and such
surface divergences promise serious difficulties.  The gravitational
implications of zero-point energy are an outstanding problem in view of our
inability to understand the origin of the cosmological constant
or dark energy \cite{straumann,weinberg89,Milton:2001np}.

As a prolegomenon to 
studying such issues, we here address a simpler question: How does the 
completely finite Casimir energy of a pair of parallel conducting plates 
couple to gravity?
The question turns out to be surprisingly less straightforward
than one might suspect!
Previous authors \cite{karim,calloni,caldwell,Sorge:2005ed,bimonte} have 
given
disparate answers, including gravitational forces,
or gravitationally modified Casimir forces,
 that depend on the
orientation of the Casimir apparatus with respect to the gravitational
field of the earth. 
  We will here resolve some of this confusion with a convincingly 
calculated result consistent with the equivalence principle.
That is, the renormalized Casimir energy couples to gravity just like any 
other energy. In our opinion, this fact is evidence that vacuum energy 
must be taken seriously in gravitational theory and that the problem of 
boundary divergences must be resolved by a better understanding of the 
modeling and renormalization processes.

We start by recalling the electromagnetic Casimir stress
tensor between a pair of parallel perfectly
conducting plates separated by a distance $a$, with transverse dimensions 
$L\gg a$, as given by Brown and Maclay \cite{bm}: 
\be
\langle T^{\mu\nu}\rangle=\frac{\mathcal{E}_c}a\, \mbox{diag} (1, -1, -1, 3),
\label{casst}
\ee
%between the plates,
where the third spatial direction is the direction normal to the plates.
This is given in terms of the Casimir energy per unit area,
$\mathcal{E}_c=-{\pi^2}\hbar c/({720 a^3})$.
Outside the plates, $\langle T^{\mu\nu}\rangle=0$.
Omitted here is a constant divergent term
that is present both between and 
outside the plates, and also in the absence of plates, which
cannot have any physical significance. Because the electromagnetic field
respects conformal symmetry, there is no surface divergent term such as is 
present for a minimally coupled scalar field subject to Dirichlet conditions
on the plates, or more generally for curved surfaces \cite{kmbook}. 
(Henceforth we will set $\hbar=c=1$.)

Now we turn to the question of the gravitational interaction of
this Casimir apparatus.
It seems to us that this question can be most simply addressed through
use of the gravitational definition of the energy-momentum tensor, as a
variation of the matter part of the action,
\be
W_g \equiv \delta W_m\equiv 
\frac12\int(dx)\, \sqrt{-g}\,\delta g_{\mu\nu}T^{\mu\nu}.\label{var}
\ee
Following Schwinger 
(note the factor of 2 in the definition),
for a weak field we define
$g_{\mu\nu}=\eta_{\mu\nu}+2h_{\mu\nu}$.
To first order we can ignore $\sqrt{-g}$.
The gravitational energy, for a static situation, is therefore given by
($\delta W\equiv -\int dt\,\delta E$)
\be
E_g\equiv \delta E_m=-\int (d\mathbf{x})\, h_{\mu\nu}T^{\mu\nu}.\label{ge}
\ee
We then replace $T^{\mu\nu}$ here by the one-loop
expectation value of the electromagnetic stress tensor
(\ref{casst}). 

\begin{figure}
\vspace{1.5in}
\includegraphics{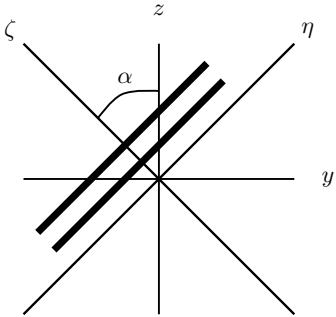}
\caption{Relation between two Cartesian coordinate frames: One attached to
the earth $(x,y,z)$, where $-z$ is the direction of gravity, and one attached
to the parallel-plate Casimir apparatus $(\zeta,\eta,\chi)$, where $\zeta$ is
in the direction normal to the plates.  The parallel plates are
indicated by the heavy lines parallel to the $\eta$ axis.  The $x=\chi$
axis is perpendicular to the page.}
\label{fig1}
\end{figure}

Calloni et al.~\cite{calloni}  and Bimonte et al.~\cite{bimonte} 
use the Fermi metric
\be
g_{00}=-\left(1+2gz\right),\quad g_{ij}=\delta_{ij}\,,\label{const}
\ee
in terms of the gravitational acceleration $g$.
This is evidently appropriate for a constant gravitational
field. We will discuss its relation to the field due to the earth below.
Let the apparatus be oriented at an angle $\alpha$ with respect to the
direction of the gravitational field.  The Cartesian coordinate
system attached to the earth is denoted by $(x, y, z)$, where, 
as noted above, $z$ is the direction of $-\mathbf{g}$.  
The Cartesian coordinates
associated with the Casimir apparatus are $(\zeta,\eta,\chi)$, where $\zeta$
is normal to the plates, and $\eta$ and $\chi$ are parallel to the plates.
(See Fig.~\ref{fig1}.)
The center of the apparatus is located at $(\zeta_0\,,\eta=0,\chi=0)$.

Now we calculate the gravitational energy from
Eq.~(\ref{ge}):
\be
E_g=\int (d\mathbf{x})\,{g z}T^{00}
=\frac{g\mathcal{E}_c}{a}\, L^2a\,\zeta_0\cos\alpha, 
\ee
up to an additive constant, independent of $\zeta_0\,$. (Any constant 
added to
the gravitational potential $g z$ simply shifts the value of this
constant.)
Thus, the gravitational force per area $A=L^2$ on the apparatus is
\be
\frac{F}A=-\,\frac1A\,\frac{\partial E_g}{\partial 
z_0}
=-g\mathcal{E}_c 
\equiv -\,\frac\epsilon{2a}\mathcal{E}_c\,,
\label{f0}
\ee
because $z_0=\zeta_0\cos\alpha$.  Note that on the earth's surface,
the dimensionless number $\epsilon\equiv 2ga/c^2$ is very small.  For a 
plate separation
of $1 \mu$m,
$\epsilon=2\times 10^{-22}$,
so the effects that we are considering are very small compared to the 
gravitational forces on the plates. (However,  Calloni et al.~\cite{calloni}
discuss the possibility of experimentally observing this force.)

It is a bit simpler to use the energy formula (\ref{ge}) to calculate the 
force by considering the variation in the gravitational energy
directly, that is,
\be
\delta E_g=-\int(d\mathbf{x})\,\delta 
h_{\mu\nu}T^{\mu\nu}.\label{1stmethod}
\ee
It is easy to verify that this gives the correct force on a mass point,
$\mathbf{F}=m\mathbf{g}$.  If we use this formula to calculate
the gravitational force on the rigid Casimir apparatus, by considering
a virtual displacement upward by an amount $\delta z_0$, we find
the same $\alpha$-independent result (\ref{f0}).

Alternatively, we can start from the definition of the gravitational
field \cite{psf},
\be
\delta W_g=\int (dx)\, \delta T^{\mu\nu} h_{\mu\nu}\,,\label{2ndmethod}
\ee
which can again be checked to yield the correct force on a mass point.
For the constant field (\ref{const}) the force on a Casimir apparatus
is obtained from the change in the energy density $T^{00}$;
that is, recalling that $z_0=\zeta_0\cos\alpha$,
\be
\delta T^{00}=\frac{\mathcal{E}_c\,\delta z_0}a\, \frac1{\cos\alpha}
[\delta (\zeta-\zeta_0-a/2)-\delta (\zeta-\zeta_0+a/2)],\label{varst}
\ee
where the $\delta$ functions arise from the step functions at the 
boundaries.
This yields from Eq.~(\ref{2ndmethod})
the same result (\ref{f0}).

Our answer is consistent with the
principle of equivalence, and  with the
second analysis of Jaekel and Reynaud \cite{jr}, who state that the
inertia of Casimir energy (at least in two dimensions) is $E_c/c^2$.
However, it is only $\frac 14$ that
found by Bimonte et al.~\cite{bimonte},
which is also the first force formula  [Eqs.\ (7) and~(8)]  provided   
by Calloni et al.~\cite{calloni}.
Our Eq.~(\ref{f0})  does, however, 
 reproduce the second formula [Eq.~(9)] given in Ref.~\cite{calloni},
which those authors describe as the one that should be observable. 
We discuss this situation further below.

We now digress to consider whether the constant-field approximation
(\ref{const}) is adequate for an apparatus suspended above the earth or 
some other pointlike mass.
  Should we instead use the perturbation of the Schwarzschild metric?
One might expect that the resulting curvature corrections are of order 
$\frac LR\ll 1$, at worst, relative to the main term, where $R$ is the
earth's radius.
The point, however, is that naive attempts to do the calculation in curved 
space change the answer by factors like $2$,
and also differ among themselves,
and the resolutions of the fallacies are sufficiently instructive to 
justify our belaboring the point.

 The Schwarzschild metric in isotropic
coordinates \cite{weinberg} and for weak fields ($GM/r\ll1$)
is
\be
ds^2=-\left(1-\frac{2GM}r\right)dt^2+\left(1+\frac{2GM}r\right)
d\mathbf{r}^2.\label{isotropic}
\ee
If we expand a short distance $z$ above the earth's surface,
we find the nonzero components of the gravitational field to be
$h_{00}=h_{11}=h_{22}=h_{33}=GM/R-gz$,
in terms of the acceleration of gravity, $g=GM/R^2$.
(It is important to recognize that the constant $GM/R$ is irrelevant in
the following, and that correspondingly the results do not depend on the
origin of $z$.)
The virtue of  isotropic coordinates is that the spatial line element
(apart from an overall factor) has the usual Cartesian form
$d\mathbf{r}^2=dx^2+dy^2+dz^2$.
Now when we compute the gravitational energy from Eq.~(\ref{1stmethod})
each component of the Casimir stress tensor contributes with equal weight:
\be
\delta E_g
=gA\,\delta z_0 \int_{\zeta_0-a/2}^{\zeta_0+a/2}d\zeta\,
(T^{00}+T^{11}+T^{22}+T^{33}),\ee
which gives the force
\be
-\,\frac1A\frac{\delta E_g}{\delta z_0}=\frac{F}A=-2ga T^{00}=
-2g \mathcal{E}_c\,,\label{r1}
\ee
since $T=T^\lambda{}_\lambda=0$.
This is twice the previous result (\ref{f0}).  
Note that again the result is independent of $\alpha$.  The same result
is obtained if we start from Eq.~(\ref{2ndmethod}).

We should be able to obtain the same result using the original Schwarzchild
coordinates, where
$h_{00}=-g z$, $h_{\rho\rho}=-g z$,
and all other components of $h_{\mu\nu}$ are zero.  However, now if we use
the first method (\ref{1stmethod}), the result is
$F/A=-4g\mathcal{E}_c\cos^2\alpha$,
so now the force depends on the orientation of the apparatus.
  Even if
$\alpha=0$, the magnitude differs from Eq.~(\ref{r1}) by an
additional factor of~2.
(It thereby fortuitously  agrees with the result in 
Ref.~\cite{bimonte} for that angle.)  

What is going on here?
The reason we get different answers in different coordinate systems
is that our starting point (\ref{ge}) is not gauge-invariant.
Under a coordinate redefinition, which for weak fields is a gauge 
transformation of $h_{\mu\nu}$ \cite{psf},
$h_{\mu\nu}\to h_{\mu\nu}+\partial_\mu\xi_\nu+\partial_\nu\xi_\mu\,$,
where $\xi_\mu$ is a vector field,  Eq.~(\ref{ge}) is invariant only if the
stress tensor is conserved, $\partial_\mu T^{\mu\nu}=0$ 
(in the weak-field context). Otherwise, there
is a change in the action,
$\Delta W_g=-2\int (dx)\xi_\nu \partial_\mu T^{\mu\nu}$.
Now in our case (where we make the finite size of the plates explicit,
but ignore edge effects  on $T^{\mu\nu}$ because $L\gg a$)
\begin{widetext}
\be
T^{\mu\nu}=\frac{\mathcal{E}_c}a\,
\mbox{diag}(1,-1,-1,3)\theta(\zeta-\zeta_0+a/2)
\theta(a/2-\zeta+\zeta_0)
\theta(\eta+L/2)\theta(L/2-\eta)\theta(\chi+L/2)\theta(L/2-\chi).
\label{explicitt}
\ee
Taking the divergence of Eq.~(\ref{explicitt}) gives corresponding $\delta$ 
functions on 
the surfaces and leads immediately to
\be
\Delta E_g=\frac{6\mathcal{E}_c}{a}\int d\eta d\chi \left[\xi_\zeta(
\zeta_0-a/2,\eta,\chi)-\xi_\zeta(\zeta_0+a/2,\eta,\chi)\right]
-\frac{2\mathcal{E}_c}a\int d\zeta d\chi \left[\xi_\eta(\zeta,
-L/2,\chi)-\xi_\eta(\zeta,L/2,\chi)\right]-(\eta\leftrightarrow\chi).\label{de}
\ee
\end{widetext}
This transformation entirely accounts algebraically for  the difference between
the forces in isotropic and Schwarzchild coordinates, but it does not yet 
explain physically why there are two different answers, nor tell us which, 
if either, is correct.

 There seem to be two possible ways to proceed. 
First, it is clear that the energy-momentum tensor of the complete 
physical situation must be conserved, and therefore the expression 
(\ref{ge}) would be gauge-invariant if we included a physical mechanism 
holding the plates apart against the Casimir attraction.
That road probably leads to complicated, model-dependent calculations.
The alternative is to
 find a physical basis for believing that
one coordinate system is more realistic than another.  Fortunately,
that problem apparently has a natural solution.
The crux of the difficulty is that the relations between coordinate 
increments and physical distances depend upon the distance from the 
gravitating center in the most common coordinate systems.

Of course, a perfect coordinate system is not possible in a curved space,
but the kind that comes closest to representing distances accurately all 
along a timelike worldline is a
 Fermi coordinate system, the general-relativistic extrapolation of
an inertial coordinate frame.  Such a system has been given by Marzlin
\cite{marzlin} for a resting observer in the field of any static mass
distribution. Here we give a simple rederivation for the case at hand.
Starting from the isotropic metric (\ref{isotropic}),  
first eliminate the constant term by rescaling the coordinates,
$t\to\left(1+\frac{GM}R\right)t$, $\mathbf{r}\to\left(1-\frac{GM}R\right)
\mathbf{r}$,
and expand to first order:
\be
ds^2=-(1+2gz)dt^2+(1-2gz)d\mathbf{r}^2.
\ee
But we need $\mathbf{r}$ to measure physical displacements even when 
$z\ne0$, so we write
\bea
&x=x'+gx'z',\qquad
y=y'+gy'z',\nonumber\\
&z=z'+\frac{g}2(z^{\prime2}-x^{\prime2}-y^{\prime2}).\label{xformfc}
\eea
Then to first order in coordinates we obtain
the Fermi metric (\ref{const}).
The corresponding gravitational force is therefore given by 
Eq.~(\ref{f0}), after all!

Now we can use the method described above to transform
the energy in isotropic coordinates to that in Fermi coordinates.  We use
Eq.~(\ref{de}) to compute the additional gravitational energy, in terms
of the gauge field $\xi_\mu$ that carries us from isotropic coordinates
to Fermi coordinates,
\be
h_{\mu\nu}^F=h_{\mu\nu}^I+\partial_\mu\xi_\nu+\partial_\nu\xi_\mu\,.
\ee
Here  
$h_{00}^I=-gz$, $h^I_{ij}=-gz\delta_{ij}\,$,
$h_{00}^F=-gz$, $h^F_{ij}=0$, $h^{I,F}_{0i}=0$.
The gauge field turns out to be 
\bea
\xi_\zeta&=&\frac12g \left(\frac12\zeta^2\cos\alpha+\zeta\eta\sin\alpha
\right)+f(\eta,\chi),\nn\\
\xi_\eta&=&\frac12g \left(\zeta\eta\cos\alpha+\frac12\eta^2\sin\alpha
\right)+g(\zeta,\chi),\nn\\
\xi_\chi&=&\frac12g \left(\zeta\cos\alpha+\eta\sin\alpha\right)\chi
+h(\zeta,\eta),
\eea
where the functions $f$, $g$, and $h$ are irrelevant.
Now from Eq.~(\ref{de}) we obtain a $\Delta E_g(z_0)$
that yields an additional force,
$\Delta F/A=g\mathcal{E}_c\,$.
When this is added to the isotropic force (\ref{r1}), we obtain the result
given in Eq.~(\ref{f0}) 
\be
\frac{F^I+\Delta F}A=-2g\mathcal{E}_c+g\mathcal{E}_c=-g\mathcal{E}_c=
\frac{F^F}A\,.
\ee

The conceptual reason why other coordinates give different answers is that 
under the virtual displacement involved in defining 
$\frac{\partial}{\partial z_0}$ one is stretching the apparatus as well as 
moving it.  Correcting for the spurious changes in $L$ and $a$ restores 
the Fermi result in all cases.
The importance of distinguishing $a$ from the physical gap was noted by 
Sorge \cite{Sorge:2005ed} in studying a related problem.

As noted above, Calloni et al.~\cite{calloni} find a result $4$ times 
ours, which is the only result from Ref.~\cite{calloni} cited in the later 
paper \cite{bimonte} with which it shares authors.
However, Ref.~\cite{calloni} states that that force formula has
two parts, in the ratio of 3/1, and that only the smaller piece 
 is  ``Newtonian,'' or ``to be tested against observation.''  
Our understanding of what that means is the following.
 Start from Eq.~(\ref{var}) and consider a general coordinate 
transformation,
$x^{\prime\mu}=x^\mu+\delta x^\mu$,
so that
$g'_{\mu\nu}(x')=g_{\mu\nu}(x)+\delta g_{\mu\nu}(x)$,
where
\be
\delta g_{\mu\nu}=\delta x^\lambda \partial_\lambda g_{\mu\nu}+g_{\alpha \nu}
\frac{\partial \delta x^\alpha}{\partial x^\mu}
+g_{\mu\beta}\frac{\partial \delta x^\beta}{\partial x^\nu}\,.\label{gct}
\ee
For a rigid translation, $\delta x^\lambda$ is a constant, so only the first
term in Eq.~(\ref{gct}) is present, which gives the result (\ref{f0}).  
However, if we do not make this restriction, we obtain from Eq.~(\ref{var})
(after integration by parts)
a surface-term correction to the force:
\be
\int_\Omega (dx)\,\sqrt{-g}f_\lambda=\frac{\delta W_m}{\delta x^\lambda}
-\int_{\partial\Omega} d\Sigma_\nu\, \sqrt{-g}T^\nu{}_\lambda\,,\label{fd}
\ee
where the force vector density is \cite{calloni} 
$-\nabla_\nu T^\nu{}_\lambda$ or
\be
\sqrt{-g}f_\lambda=
-\partial_\nu\left(\sqrt{-g}T^\nu{}_\lambda\right)
+\frac12\sqrt{-g}T^{\mu\nu}\partial_\lambda g_{\mu\nu}\,. \label{lambda}
\ee
Note that the surface term
identically cancels the first term in $f_\lambda\,$.  
Now if $\Omega$ refers to all space, the surface term vanishes (as we 
have shown explicitly).
But if $\Omega$ is just
the space volume  between the plates,
and we include this correction for the Fermi metric (\ref{const}) 
for which $\sqrt{-g}=1+g z$, we obtain an additional term
$-3g\mathcal{E}_c\cos\alpha$. 
Adding this to the previous result (\ref{f0}), 
we obtain the result of Ref.~\cite{bimonte}
if  $\alpha=0$. However, in general the result depends on the angle
between the apparatus and the vertical.
 Is this consistent with the equivalence principle 
 (the scalar nature of mass)?
A similar angle dependence will now occur with the isotropic
Schwarzchild metric.  Why should one trust the formula (\ref{lambda}) 
over
the more fundamental variational principle when boundaries are present?
Omission of the surface term resolves the discrepancy, giving the 
equivalence-principle result (\ref{f0}).

After circulation of the present paper as arXiv:hep-th/0702091, Bimonte et
al.~\cite{bimonte2} responded that (i) our result is implicit in their earlier 
analysis; (ii) our Eq.~(\ref{var}) implicitly assumes the equivalence 
principle, which they identify with the conservation law 
$\nabla_\mu T^{\mu\nu}=0$. 
We reply:  (i) Ref.~\cite{bimonte} states without qualification that the force 
is 4 times the correct value.  
(ii) By ``equivalence principle'' we merely mean 
that gravity tends to enforce geodesic motion.  
This is not obviously equivalent
to conservation, as shown by the existing uncertainty in the literature.  
Contrary
to statements in Ref.~\cite{bimonte2}, it is generally accepted in quantum 
field theory in curved space-time that the renormalized total $T^{\mu\nu}$ 
must be conserved.

\begin{acknowledgments}
This work is supported in part by Collaborative Research Grants from
the US National Science Foundation and in part by the
US Department of Energy. It was completed while S.A.F. enjoyed 
 partial support at the I. Newton Institute for 
Mathematical Sciences, Cambridge. We thank Gabriel Barton, Michael Bordag,
Robert Jaffe, and Ron Kantowski for helpful conversations and emails.

\end{acknowledgments}

\end{document}